\documentclass[useAMS,usenatbib]{mn2e}
\usepackage{psfig}
\usepackage{graphicx}
\usepackage{amsmath}
\usepackage{amssymb}

\newcommand{\adsurl}[1]{}
\providecommand{\url}[1]{{#1}}

\newcommand{\be}{\begin{equation}}
\newcommand{\ee}{\end{equation}}
\newcommand{\bea}{\begin{eqnarray}}
\newcommand{\eea}{\end{eqnarray}}
\newcommand{\bn}{{\mathbf n}}
\newcommand{\bd}{{\mathbf d}}
\newcommand{\bq}{{\mathbf q}}
\newcommand{\clean}{\texttt{clean} }
\newcommand{\cleanly}{\texttt{clean}ly }
\newcommand{\cleanbc}{\texttt{clean$_{bc}$} }
\newcommand{\sclean}{\texttt{superclean} }
\newcommand{\scleanbc}{\texttt{superclean$_{bc}$} }
\makeatletter
   \newcommand\tabcaption{\def\@captype{table}\caption}
\makeatother

\title[Galaxy Zoo: The large-scale spin statistics of spiral galaxies in the Sloan Digital Sky Survey.]
{Galaxy Zoo: The large-scale spin statistics of spiral galaxies in the Sloan Digital Sky Survey
\thanks{This publication has been made possible by the participation of more 
than 100,000 volunteers in the Galaxy Zoo project;
www.galaxyzoo.org/Volunteers.aspx}}
\author[Kate Land et al.]
{Kate Land$^1$\thanks{E-mail: krl@astro.ox.ac.uk}, 
An\v{z}e Slosar$^{1,2}$\thanks{E-mail: anze@berkeley.edu}, 
Chris Lintott$^1$, 
Dan Andreescu$^3$, 
Steven Bamford$^4$, \cr 
Phil Murray$^5$, 
Robert Nichol$^4$, 
M.Jordan Raddick$^6$, 
Kevin Schawinski$^1$, \cr 
Alex Szalay$^6$, 
Daniel Thomas$^4$, 
Jan Vandenberg$^6$.\\
\vspace*{-6pt} {\small \em $^1$Astrophysics, University of Oxford, Denys Wilkinson Building, Keble Road, Oxford, OX1 3RH, UK}\\
\vspace*{-6pt} {\small \em$^2$ Berkeley Center for Cosmological Physics, Lawrence Berkeley National Laboratory and Physics  Department, University of California,}\\\vspace*{-6pt}{\small \em Berkeley CA 94720, USA}\\
\vspace*{-6pt} {\small \em$^3$ LinkLab, 4506 Graystone Ave., Bronx, NY 10471, USA}\\
\vspace*{-6pt} {\small \em$^4$ ICG, University of Portsmouth, Mercantile House, Hampshire Terrace, Portsmouth, PO1 2EG, UK}\\
\vspace*{-6pt} {\small \em$^5$ Fingerprint Digital Media, 9 Victoria Close, Newtownards, Co. Down, Northern Ireland, BT23 7GY, UK}\\
\vspace*{-6pt} {\small \em$^6$ Department of Physics and Astronomy, The Johns Hopkins University,
Homewood Campus, Baltimore, MD 21218, USA}}

\begin{document}

\date{Accepted xxx. Received xxx; in original form xxx}

\pagerange{\pageref{firstpage}--\pageref{lastpage}} \pubyear{2008}

\maketitle

\label{firstpage}


\begin{abstract}
  We re-examine the evidence for a violation of large-scale
  statistical isotropy in the distribution of projected spin vectors
  of spiral galaxies. We have a sample of $\sim 37,000$ spiral
  galaxies from the Sloan Digital Sky Survey, with their line of sight
  spin direction confidently classified by members of the public
  through the online project Galaxy Zoo~\citep{GZ}.  After
  establishing and correcting for a certain level of bias in our
  handedness results we find the winding sense of the galaxies to be
  consistent with statistical isotropy.  In particular we find no
  significant dipole signal, and thus no evidence for overall
  preferred handedness of the Universe.  We compare this result to
  those of other authors and conclude that these may
  also be affected and explained by a bias effect.
\end{abstract}



\section{Introduction}

\begin{figure*}
\begin{minipage}{8.5cm}
\centering
\includegraphics[width=2cm]{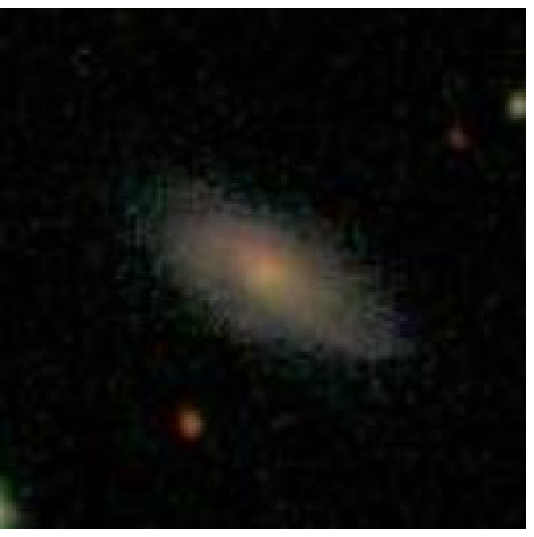}
\includegraphics[width=2cm]{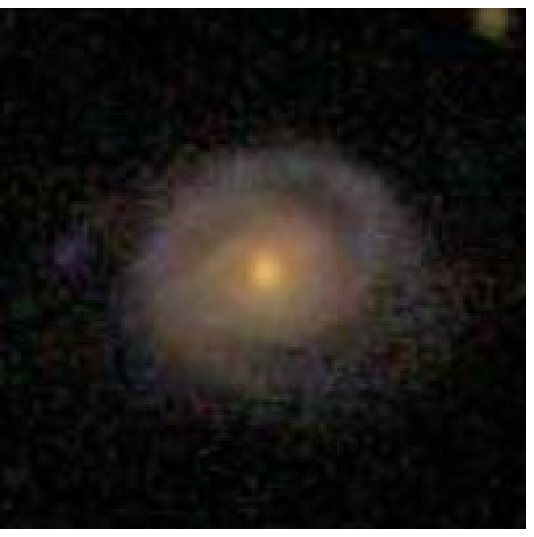}
\includegraphics[width=2cm]{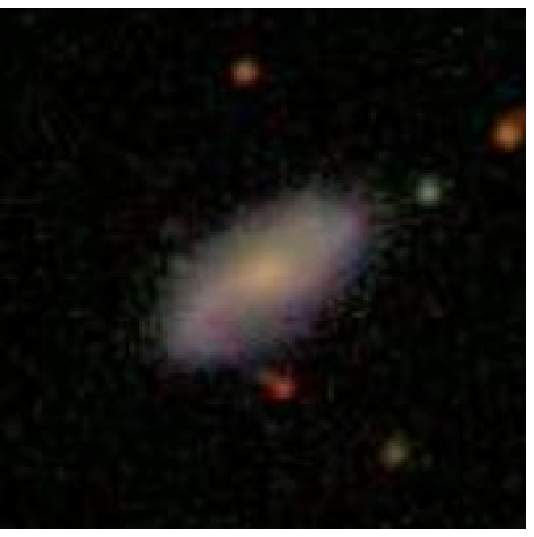}
\includegraphics[width=2cm]{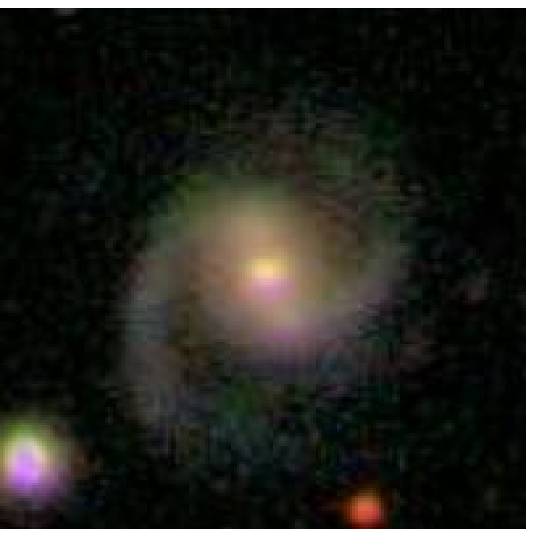}\\
\vspace{2pt}
\includegraphics[width=2cm]{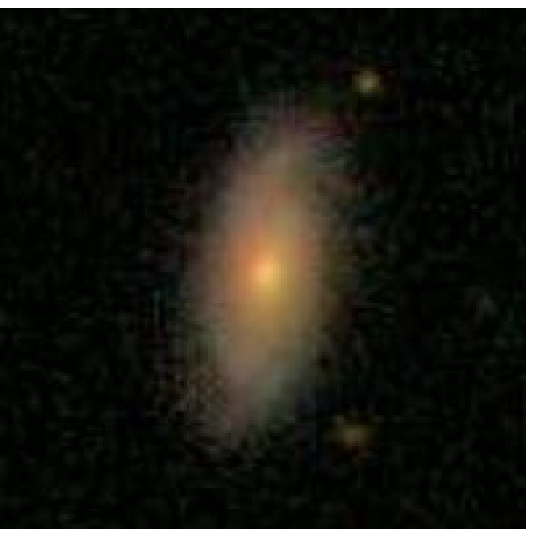}
\includegraphics[width=2cm]{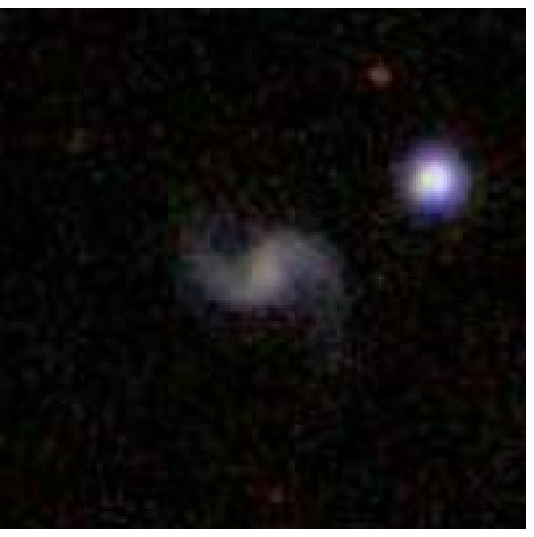}
\includegraphics[width=2cm]{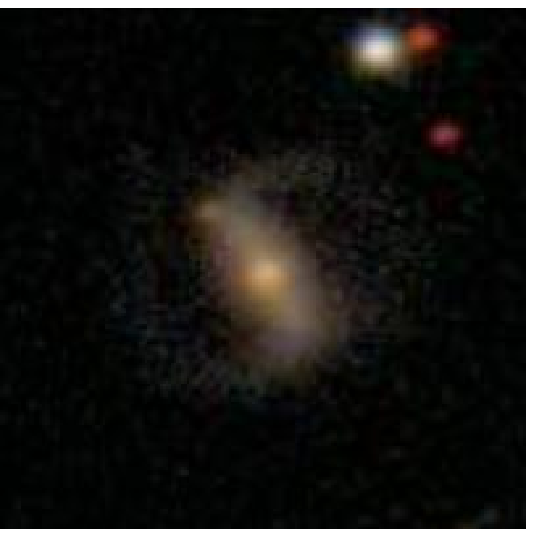}
\includegraphics[width=2cm]{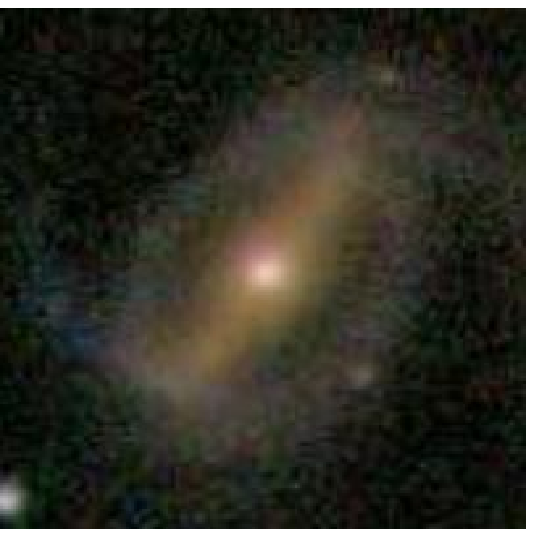}
\caption{Typical Z and S-wise galaxies in our \clean sample. 
The top four images are classified as Z-wise (with class-weights of 
85.0\%, 88.2\%, 90.6\%, 94.6\%), and the bottom four as S-wise 
(with class-weights of 
84.0\%, 86.2\%, 87.3\%, 93.9\% from left to right).}\label{eg}
\end{minipage}\hfill
\begin{minipage}{8.5cm}
\begin{tabular}{ccl}
\hline
Class & Button & Description \\
\hline
1 & \raisebox{-0.5ex}{\includegraphics[width=0.5cm]{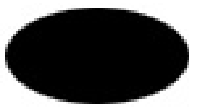}} 
& Elliptical galaxy \\
2 & \raisebox{-2.0ex}{\includegraphics[width=0.5cm]{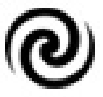}}
 & Clockwise/Z-wise spiral galaxy \\
3 & \raisebox{-2.0ex}{\includegraphics[width=0.5cm]{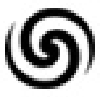}}
 & Anti-clockwise/S-wise spiral galaxy \\
4 & \raisebox{-2.0ex}{\includegraphics[width=0.5cm]{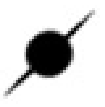}}
 & Spiral galaxy other (e.g. edge on, unsure)\\
5 & \raisebox{-2ex}{\includegraphics[width=0.5cm]{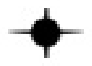}} 
& Star or Don't Know (e.g. artefact) \\
6 & \raisebox{-2ex}{\includegraphics[width=0.5cm]{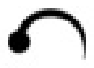}} 
& Merger \\
\hline
\end{tabular}\tabcaption{The Galaxy Zoo classification scheme. The symbols are the 
same as those used on the buttons of the Galaxy Zoo analysis web page. Note that 
spiral galaxies have three separate class-types, while ellipticals have just one.
}\label{class}
\end{minipage}
\end{figure*}

Galaxy Zoo\footnote{www.galaxyzoo.org}~\citep{GZ} is a online project
in which volunteers visually classify the morphologies of galaxies
selected at random from the spectroscopic sample of the Sloan Digital
Sky Survey (SDSS,~\cite{York}) Data Release 6 (DR6). Information of
this type is essential to the development of our understanding of
galaxy formation.  The public response to the launch of Galaxy Zoo in
July 2007 was overwhelming, achieving over 36 
million classifications within a few months and results that agree exceptionally well
with those of professional astronomers~\citep{GZ}. To make the project
accessible to as many people as possible a relatively simple
classification scheme was used, outlined in Table~\ref{class}.

Along with the morphology of the galaxies the perceived rotation
direction was also requested where possible, i.e. for spiral galaxies,
where it is assumed that the arms of the spiral galaxies trail the
rotation. Determining the rotation
direction in this way is accurate for 96\% of galaxies~\citep{pasha}. 
Moving from the centre of the galaxy outwards,
the spiral in fact `unwinds' the opposite way to the rotation, and
therefore to avoid further confusion we herein refer to a clockwise
and anti-clockwise classification as Z and S-wise respectively,
indicating the projected pattern of the galaxy arms on the sky. A
Z-wise galaxy (see Table~\ref{class}) is assumed to be rotating
clockwise with respect to our line of sight, thus with its angular
momentum pointing away from us.  Likewise, an S-wise galaxy is assumed
to be rotating anti-clockwise (also known as `counterclockwise'),
with its angular momentum vector pointing towards us.

We perform a large-scale multipole analysis of the Galaxy Zoo
catalogue of spiral galaxies winding sense, accounting for the partial
sky coverage and quantifying the level of uncertainty introduced by
this.  Locally ($|cz| \lesssim 3000$ km s$^{-1}$) this information can
be used to constrain galaxy formation scenarios~\citep{SZwise}, while
at higher redshifts we can test the fundamental assumption that the
Universe is statistically homogeneous and isotropic over cosmological
scales.  In a companion paper~\citep{2point} we use the same spin data
to analyse the small-scale 2-point correlation function, relating the
results to predictions from the tidal-torque theory of
structure-momentum formation and N-body simulations.

In Section~\ref{data} we introduce the Galaxy Zoo (GZ) dataset, and
basic classification results.  Due to the high signal-to-noise (i.e. multiple classifications)
of our dataset, we are able to further investigate the level of bias
that may be introduced into our handedness results because of the
visual and human nature of the classifications, and in
Section~\ref{bias} we discuss these results and present the final
dataset in Section~\ref{finaldata} that we use thereafter.  In
Section~\ref{analysis} we outline our method of analysis, and the
results for our GZ dataset. We also consider third-party datasets in
Section~\ref{third}, and provide complementary results to those
of~\cite{Longo, SZwise}, with some clarifying comparisons. We discuss
our results in Section~\ref{end}.


\section{The Data}\label{data}

As of November 28$^{\rm th}$ 2007, GZ had over $36$ million
classifications (called 'votes' herein) for 893,212 galaxies from
85,276 users. This sample of galaxies were selected to 
be those that are targeted for spectroscopy by SDSS; 
extended sources with Petrosian magnitude $\texttt{petroMag\_r} < 17.77$. We 
also included objects that were not originally targeted as such, but were 
observed to be galaxies once their spectrum was taken.
Where spectroscopic redshifts are available, we find they range to $z \lesssim 0.5$, 
with an average of $\bar{z} \sim 0.14$ ($\sim 600$ Mpc). The galaxies
thus probe our local universe at cosmological scales.
Every galaxy in this SDSS Data Release 6 (DR6)
spectroscopic sample has been classified on average $\sim 41$ times.
In practice $\sim 5\%$ of this raw data is removed when we only allow
one vote per galaxy per person, taking the first vote where necessary
(multiple votes are probably due to people double clicking with the
mouse). This results in an average of $\sim 39$ distinct votes per
galaxy.

To reduce this information to one final classification (and
corresponding uncertainty) per galaxy a variety of different schemes
were investigated.  Firstly, for each galaxy we simply counted the
number of votes it received for each of the six class-types (see
Table~\ref{class} for an explanation of the six different
class-types), resulting in six class-weights for each object that sum
to 1 (or 100\%).  The largest class-weight then indicates the final
classification of the object, although in practice we only use objects
where one class-type receives a significant majority of the
votes. This scheme is called `unweighted' as each vote is counted
equally in computing the class-weights, regardless of the user.

An alternative method involved weighting the raw votes such that users
who tended to agree with the majority have their votes up-weighted,
while users who consistently disagreed with the majority have their
votes down-weighted. The user-weights are iteratively determined by
considering for each object (at each iteration) how well a user's vote
agrees with the class-weights for that object, as determined from the
weighted votes from \emph{all} users.  All users start with a weight
of 1, but after a number of interations this converges, with the final
user-weights used to establish the final class-weights for each
object.
This method has the advantage of down-weighting users who are
consistently unreliable. However it assumes that the majority vote is
always right, and thus may penalise users who are more careful than
the average user.

In the end we find that the results do not differ significantly
between the weighted and unweighted
schemes. 
Therefore we herein use the unweighted results as their simplicity (in
that they do not correlate the data across the galaxies) will be
essential in Section~\ref{bias} when we consider subsets of our data
in the analysis of possible bias effects in our results.

From our GZ catalogue of final class-weights we identify the objects
with relatively low final classification uncertainty.  Our \clean and
\sclean samples contain objects with at least 10 votes (which is
almost all of our GZ galaxies), and a top class-weight of over 80\%
and 95\% respectively.  This guarantees at least a 5$\sigma$ and
7$\sigma$ detection of the final classification respectively (assuming
people were voting at random from the six options), and in most cases
(i.e. for 39 votes) at least 10$\sigma$ and 13$\sigma$
respectively. Of course, these significances are purely statistical
and there is some human `systematic' error involved that is hard to
quantify due to the nature of the experiment. We are justified in
claiming, however, that more votes would not change our sample beyond
noise fluctuations as the votes are uncorrelated.

Our final \clean sample contains 291,626 objects (when class=2,3,4
votes are all counted as `spiral'), and thus we have the morphological
classifications at over $5 \sigma$ confidence for approximately a
third of the entire spectroscopic sample of SDSS DR6.  Of these we
find 97,848 to be spiral galaxies and 184,743 are elliptical galaxies,
with the remaining as stars (8,074) and mergers (961). We find the
galaxy morphology classifications agree exceptionally well with those
of expert classifications such as~\cite{Fuk} and~\cite{Kev}
(see~\cite{GZ} for more details).

\begin{figure}
\centerline{\includegraphics[width=7cm]{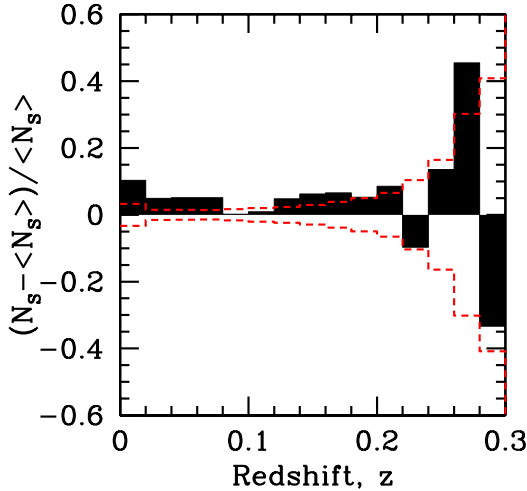}}
\caption{The S-wise excess as a function of redshift. In redshift bins of 
$\Delta z = 0.02$ we count the number of S-wise galaxies that we find, $N_S$, and the 
number we would expect assuming that S and Z-wise galaxies are equally likely, i.e. 
$\left<N_S\right>=N/2$ where $N$ is the total number of galaxies in the redshift bin. 
We plot the difference between the observed and expected numbers, 
normalised by $\left<N_S\right>$ just for visual clarity. We also show the 1$\sigma$ 
scatter (red dashed histogram) expected assuming the normal approximation to 
binomial statistics, $\sigma = 1/\sqrt{N}$. }\label{histo}
\end{figure}

\subsection{The Initial Spin Sample}

In this paper we are interested in the projected spin classifications of the spiral 
galaxies; the galaxies classified as class=2 or class=3. We 
find (17,100, 18,471) of the spiral galaxies are \cleanly 
classified (i.e. over 80\% weights) with (Z, S)-wise winding sense respectively, 
with equivalent (Z, S)-wise number counts of (6,106 , 7,034) for the \sclean sample. 
For a null hypothesis of statistical isotropy we 
would expect (Z, S)-wise handedness to be equally likely across the sky, however 
we observe a significant excess of S-wise galaxies in 
both our samples, at more than $7 \sigma$ significance\footnote{As 
determined from the normal approximation to the binomial distribution valid for 
large sample sizes, which indicates that the number of S-wise galaxies out 
of a total of $N$ galaxies has a probability distribution 
$\sim {\mathcal N}(N/2,N/4)$.}.

We consider how this S-wise excess varies with redshift, to gain
insight into the possible source of this signal. Of the SDSS galaxies 
targeted for spectroscopy, $\sim70\%$ have had their 
spectrum taken and processed in DR6, and we find (12,055, 13,123) of our \clean (Z,
S)-wise galaxies have spectroscopic redshifts available.  From these
we plot a histogram of the S-wise excess as a function of redshift in
Figure~\ref{histo}. In each redshift bin we find the total number of
galaxies $N_i$, and the expected number of S-wise galaxies assuming
statistical isotropy, i.e. $\left<N_s\right>=N_i/2$. We then plot the difference
between the number of S-wise galaxies observed in this bin, $N_s$, and
the expected number. We further normalise by the expected number for
clarity. We also include the 1$\sigma$ error bars from assuming a
normal approximation to the binomial distribution. We see that the
excess is fairly consistent across our sample, and not associated with
any particular redshift. Different physical processes dominate at
different redshifts, and thus the consistency of this S-wise excess is
perhaps indicative of some overall bias in our results rather than a
true astronomical signal.


We investigate this possibility further in the next section, but as a
first check we consider the full $\sim 35$ million votes that
contribute to our final classifications. We find 2,272,354 votes for
Z-wise and 2,482,271 for S-wise.  Compared to the total number of
votes, and total number of objects, this roughly indicates that there
are 58,632 Z-wise and 64,049 S-wise galaxies in our GZ dataset, and an
excess of S-wise galaxies at over 15$\sigma$.  The S-wise proportion
from the raw-vote counts is 52.5\%, which is higher than the 51.9\%
returned from the \clean spin catalogue number counts. This possibly
indicates the presence of some biasing effect, as the raw vote
proportions should match the \clean number-count proportions if there
is no bias, as these should both just reflect the true proportions in
our sample.



\subsection{The Bias Study}\label{bias}

Before we analyse our data we wish to check that our GZ \clean spin catalogue 
(those galaxies with final \clean classifications of class=2 or class=3) 
provides a fair representation of the full GZ dataset. It is important that we make 
these checks as there are a number possible sources of bias in the GZ 
classification scheme.
The process of visually classifying the winding sense of galaxies may 
introduce a bias if humans are able to discern patterns of one handedness 
better than others - an effect that some neuroscientists believe to 
exist~\citep{neuro}. The design of the Galaxy Zoo website may also be 
influencing the decisions of the users, through the symbols on the buttons or 
the images shown in the tutorial. Alternatively, the 
position of the buttons on the screen could cause users to 
sometimes record their votes incorrectly, in a systematic way. 

It could be argued that there may exist detectable malicious contamination from a 
small number of users recording incorrect classifications, in a systematic or random way. 
However, we only allow one vote per person 
per galaxy and so the effect of individual users is very limited. Further, with regard to the 
S-wise excess, if we consider the number of class=2 and 
class=3 votes that each user records then we observe an \emph{overall trend} 
in the results - with all users generally clicking the S-wise button more 
then the Z-wise button. Thus we know that the S-wise excess is not just coming 
from the results of a few users.

Before we introduce the results of our bias study, let us just clarify how 
a biasing effect could influence our spin catalogue. We do not hypothesise that 
any biasing effect could be so large that 
galaxies in our \clean sample are misclassified, i.e. they have been assigned an 
incorrect final classification. This 
would require $\gtrsim 35$ different people who do not know each other to all 
select the \emph{same} incorrect button when they view that galaxy.
Instead, a biasing effect would impact our spin catalogue if it makes it harder for 
some types of galaxies to be \cleanly classified than other types. Take 
for example a human bias effect, in which a human can discern a S-wise spiral pattern 
better than a Z-wise spiral pattern. This would mean that a S-wise galaxy has 
slightly more chance of being in our \clean sample than a Z-wise galaxy, resulting 
in perhaps more S-wise than Z-wise galaxies in our \clean sample even if there are an 
equal number on the sky.

\begin{table}
\begin{centering}
\begin{tabular}{c|rl|rl}
\hline
 class & monochrome & & mirrored & \\
 -type & $<\%> $ & $\sigma$& $<\%> $ & $\sigma$\\
\hline
           1 & 55.962  & 0.123  &55.015   &0.124 \\
           2 & 5.525  & 0.071   &5.646   &0.071 \\
           3 & 6.032  & 0.073  & 5.942  & 0.072 \\
           4 & 17.416  & 0.092  &18.461 &  0.093 \\
           5 & 11.059  & 0.060 & 11.265  & 0.060 \\
           6 & 4.007  & 0.050  & 3.670   &0.045 \\
\hline
\end{tabular}
\tabcaption{The average class-weights as a function of class-type for 
the monochrome data and mirrored data. 
We see that the average class=2,3  weights do not reverse between the monochrome and 
mirrored images, therefore indicating that there is a bias effect in the data. The 
evidence for a true excess of S-wise of galaxies on the sky is thus limited.}\label{biastab}
\end{centering}
\end{table}

To investigate the possibility of bias-effects being present in our spin catalogue 
we obtain classifications for mirror images of a subset of the GZ dataset. This 
will conclusively constrain the level of bias effecting our spin results, although 
we will not necessarily be able to identify the source without further investigations.
Extra images of the entire \sclean sample (as of 4$^{\rm th}$ Sept 2007) and a 
random 5\% of the rest of the GZ catalogue were used. This `bias sample' 
contained 91,303 objects, and for each we submitted to the 
GZ website 3 transformed images: a monochrome version, 
a vertically mirrored version, and a diagonally mirrored version.
From the 28$^{\rm th}$ November 2007 
it has been this bias sample that the website has been collecting classifications 
for, and as of January 5$^{\rm th}$ we had an average of 22 distinct votes per image 
that we use herein.

We condense these votes into final classifications as before. We observe 
effectively no difference in the results from the two different mirror images, and so 
we combine the mirror votes. Therefore, for each of the objects in our bias 
sample we have 
two new sets of class-weights - one set from the monochrome image, 
and the other from the mirror images. In comparing these two new class-weights 
we can assess the level of any bias in our results. 
We cannot compare these class-weights with those of 
the original images, 
as the votes were logged at different times and we find the behaviour 
of users to vary at a detectable level from month to month (this is 
especially true when a newsletter is sent out to the users and the 
site receives a surge of traffic). Only votes that 
have been obtained concurrently can be compared completely fairly.

We perform two different tests for bias in our handedness results (see~\cite{GZ} for an 
analysis of the bias results with respect to the elliptical vs spiral classifications). 
Firstly we consider the 
average class-weight for class=2 and 3, and examine how this 
changes between the monochrome and mirrored images. 
We obviously expect to see the average class-weight 
for class=3 to be higher than class=2 for the monochrome images to reflect the 
S-wise excess that we observe. We then also expect to see this average class-weights  to 
switch over for the mirrored images.

In Table~\ref{biastab} we record the average class-weight (as a percentage) that 
each class-type receives for the monochrome and mirrored 
images~\footnote{This is actually not just a simple average over our bias sample, 
as not all of the galaxies in the bias sample were selected at random.
We up-weight the results of the randomly selected portion of our bias sample 
to balance with the \sclean portion and effectively create a random subsample of the 
full GZ dataset. We find the same results if we down-weight the \sclean 
portion of our bias sample, or select just 5\% of them at random.}. 
We obtain estimates for the errors on our averages through the jackknife method of 
resampling (with 2000 samples). 



We find that the average (Z, S)-wise weights are (5.5\%, 6.0\%) for
the monochrome images.  The significantly higher S-wise average weight
(at $\sim 5\sigma$) agrees with our observation of an S-wise excess of
galaxies.  If there is no bias in our handedness results then we would
expect these class-weight averages for the mirror images to swap over,
with now a higher class-weight for Z-wise.  However, we find average
class-weights of (5.6\%, 5.9\%) for the mirrored images - which still
displays a significantly higher S-wise average weight, at $\sim 3
\sigma$.  The fact that the weights stay the same within statistical
accuracy indicates that we have a significant level of bias in our
results, and no true S-wise excess. In particular, we can put an upper
limit on the intrinsic excess of S-wise galaxies to $|N_{\rm
  S}-N_{\rm Z}|/(N_{\rm S}+N_{\rm Z})<0.021 (0.028)$ at 95\% (99.7\%)
confidence limits.


Secondly, we assess the level of bias that may be effecting the \clean
and \sclean samples.  We expect these samples to contain relatively
visually clear galaxies, and therefore it is not obvious that a subtle
effect such as that found for the average GZ galaxy would still effect
these samples. We consider an effectively random sample of GZ
galaxies~\footnote{We use the random portion of our bias sample, and a
  number of the \sclean portion selected at random with the same 5\% 
  probability.}. We consider the number of these galaxies that
pass the classification criteria of the \clean sample for the
monochrome images, and for the mirrored images.  We find (Z, S)-wise
numbers of (839, 923) for the monochrome images, and (864, 905) for the
mirrored images. Therefore we again see the bias effect, as they both
find more S-wise galaxies (although due to the smaller sample size the
effect is not as significant as before, with jackknife errors of $\sim 25$ galaxies). We can 
however always eliminate any possible bias effect by insisting that an image passes the criteria
before AND after mirroring, and in this case we find (739, 739) (Z,
S)-wise galaxies.  This indicates that the bias effect can account for
the excess exactly.


\begin{figure*}
\centerline{\includegraphics[angle=90, width=8.cm]{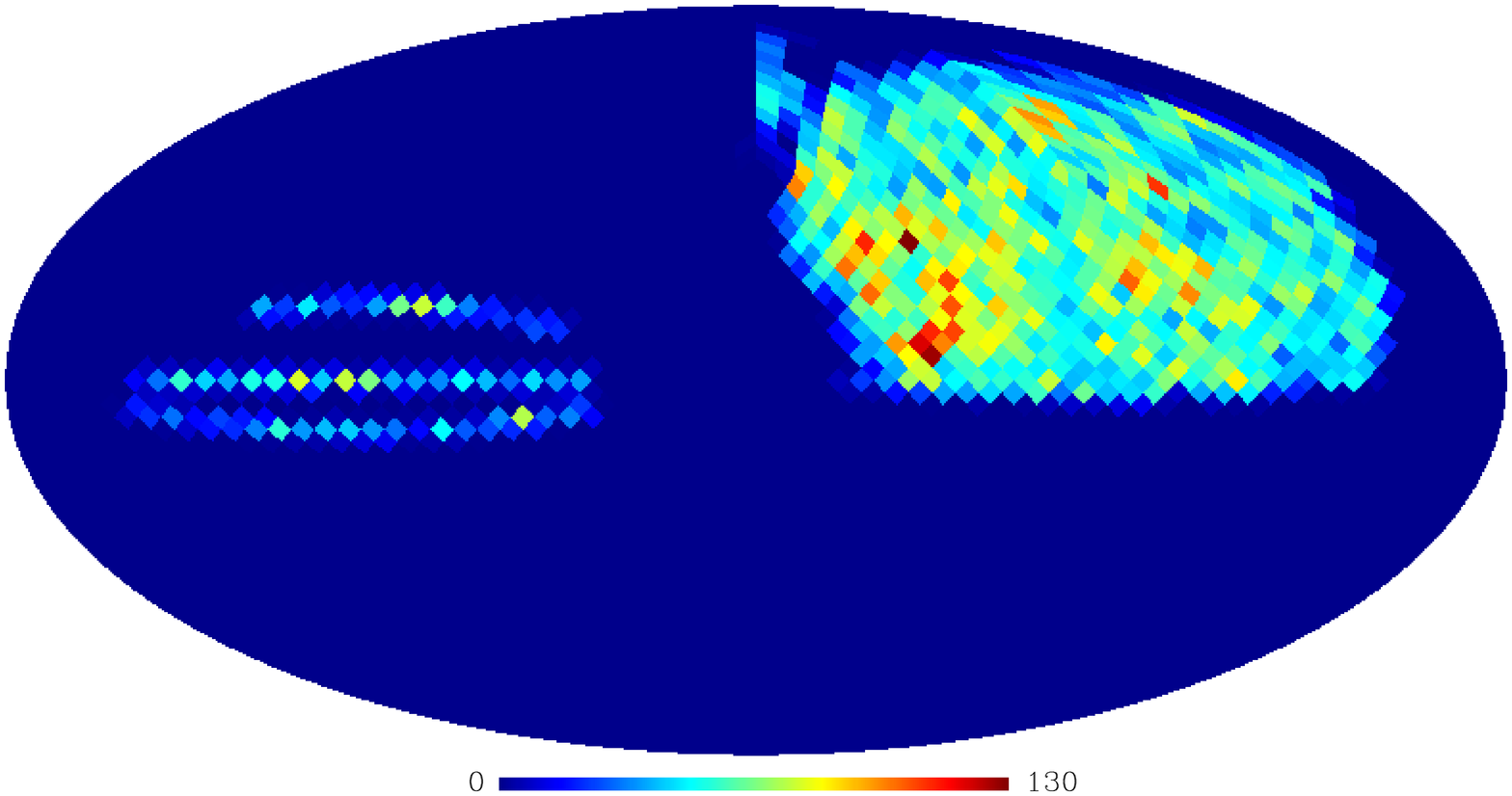}
\includegraphics[angle=90, width=8.cm]{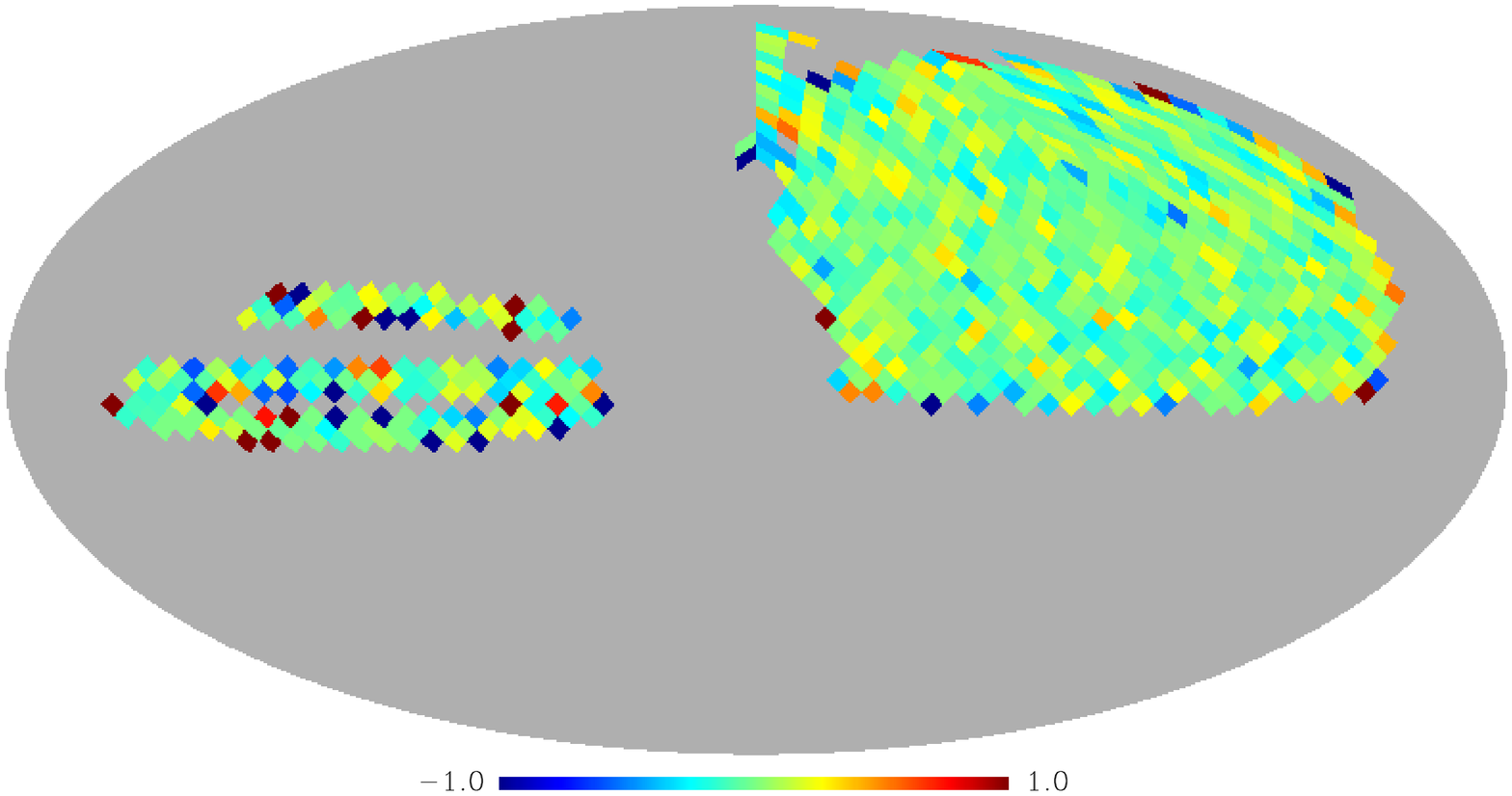}}
\caption{The number of galaxies per pixel (left) and their average spin 
(right) for our \clean galaxies. Plotted in elliptical (Mollweide) projection 
with the Healpix pixelisation scheme, with the Equatorial coordinate 
system rotated to centre on (RA,DEC)=$(-90^\circ,0^\circ)$.}\label{sky}
\end{figure*}

We find it is beyond the scope of this paper to identify the bias
source.  It may be that people find it easier to `see' the handedness
of an S-wise galaxy. However, we cannot tell the difference between
`not seeing' the handedness and clicking the wrong button (by
accident). So the two main culprits remain the design of the site, and a
human pattern recognition effect.

The latter case would be of interest to neuroscientists. For example, 
in~\cite{neuro} the authors found that observers who stared 
at the centre of a Leviant's `Enigma' illusion (see their Figure 1) 
would judge the perceived rotation as clockwise for longer 
than anti-clockwise (as the motion 
changes direction)~\footnote{We thank Jim Grange for these interesting 
comments, and directing us to this work.}. No explanation for the directional bias 
was found, but this effect might cause GZ users to be biased in any motion 
that they perceive in unclear images 
of spiral galaxies. Whether this leads to a greater chance of selecting 
the S-wise icon, or influencing the results for relatively clear images is 
not evident.


\subsection{The Final (bias corrected) Dataset}\label{finaldata}

The bias study results have demonstrated that there is not an excess of S-wise galaxies 
at any significant level, but rather that the S-wise class on average 
receives extra votes for some reason - making it a bit easier for a S-wise galaxy to
pass the \clean criteria than a Z-wise galaxy.  
Individual \clean classifications are correct, but overall number counts are skewed and 
we would like to correct for this effect.
We do so by reducing the classification criteria 
for the Z-wise class-weight.

We find Z-wise \clean and \sclean classification 
criteria of 0.78 and 0.94 respectively brings the numbers of S and Z-wise galaxies into 
agreement within $1\sigma$, and we employ this Z-wise criteria herein. However, in all 
our statistical analysis we will marginalise over this correction, 
and therefore fold in all the uncertainty due to the bias effect.

This is a rather crude bias correction, as the level of bias may vary with 
redshift, magnitude, size, etc. like that of the morphology bias examined in~\cite{Bamford}. 
However, we do not know how the true Z/S -wise ratio varies with distance or magnitude, 
and thus we cannot make a bias correction that varies with such parameters. Therefore 
Z-wise and S-wise number counts cannot in general be compared for a subset of our 
`bias corrected' dataset.

We have a new bias corrected (`bc') \cleanbc sample of (Z,S)=(18,467, 18,471) and a 
\scleanbc sample of (7,047, 7,034). 
From this we remove probable duplicate objects which 
can arise when a large galaxy is given more than one spectroscopic ID 
during the SDSS pipeline. We use 
the maximum \texttt{petroRad\_r} of a pair to determine if their 
\texttt{Objid}s actually point to the same object, and remove one of them at random, 
resulting in a \cleanbc spin sample of (Z,S)=(18,074, 18,052) galaxies, and 
similarly a \scleanbc sample of (Z,S)=(6,902, 6,894).
For visualization, we reduce this data into a pixelised map 
by averaging the spins of the galaxies
\be
\frac{\sum_{j=1}^{N_i} s_j}{N_i}
\ee
where the summation is over the $N_i$ galaxies in the $i$th pixel, and 
$s_j=+1,-1$ if the galaxy is Z, S-wise respectively.
We use the Healpix pixelisation scheme~\citep{healp}, with nside=16. 
The number of pixels in our map$=12\:\textrm{nside}^2=3,072$, of 
which our \cleanbc spin sample covers 801, and we plot these results in Figure~\ref{sky}.

\section{Analysis}\label{analysis}


\subsection{Method}\label{anal_method}
We wish to establish the large scale statistical properties of the galaxy spins. 
Although there is some level of uncertainty in the overall (S, Z)-wise number counts, it is still 
possible to look for a dipole, for example, in the spin distributions.
Rather than 
using an averaged map, such as that in Figure~\ref{sky} we fit a probability model to all 
of the galaxies. The null hypothesis that we wish to test against is that we are 
equally likely to observe an Z or S-wise galaxies wherever you look on the sky - i.e. 
statistical isotropy. However, we wish to constrain alternative models in which the 
probability of observing a Z or S-wise galaxy varies with position on the sky (
the sum of these probabilities must always equal one). To explore this 
possibility we expand the probability of 
a galaxy spin being Z-wise into the first few (the largest) spherical harmonic 
modes, such that it is a function of position on the sky;
\bea
P(Z | \hat{\bn}) &=& 0.5 + M + D\:\hat{\bd}.\hat{\bn} + Q\:(\hat{\bq}_1.\hat{\bn}\:\hat{\bq}_2.\hat{\bn}-\frac{1}{3}\hat{\bq}_1.\hat{\bq}_2)\nonumber\\
P(S | \hat{\bn})&=&1-P(Z | \hat{\bn})\label{mod}.
\eea
where $M, D, Q$ are the magnitude of the monopole, dipole, and quadrupole 
respectively.

This parameterisation of harmonic modes in terms of unit vectors, and an 
overall magnitude, is known as Maxwell 
Multipole Vector representation, and it is increasingly being 
used in CMB analysis~\citep{copiMV, MV}. 
Compared to the standard expansion into spherical harmonic coefficients it 
has the advantage of providing rotationally invariant parameters, or 
parameters that rotate simply with the coordinates (such as the vectors). 
The vectors are in fact `headless' - a change in sign can be absorbed by the 
magnitude $D, Q$. To avoid this degeneracy we restrict the magnitude of $D$ and $Q$ 
to be positive. For the quadrupole the vectors can still both flip signs together, but 
in reality we find this degeneracy to not be a problem as the 
MCMC methods tend to converge to one solution.

We wish to explore the constraints that our Galaxy Zoo dataset provides for 
these parameters. We can find the probability of the parameters 
from the likelihood of our data
\be
L = \prod_i P(h_i | \hat{\bn}_i )\label{like}
\ee
where $h_i$ is the handedness of the $i$th galaxy (Z or S), which is at position 
$\hat{\bn}_i$ on the sky.


\subsection{Results}\label{results}

\begin{figure}
\centerline{\includegraphics[width=8.5cm]{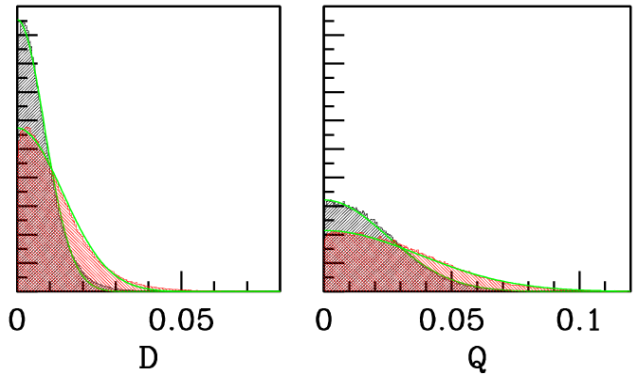}}
\caption{The marginalised constraints on the dipole $D$ and Quadrupole $Q$ as defined 
in (\ref{mod}), from MCMC analysis of the likelihood (\ref{like}). Results are 
shown for the \cleanbc (grey) and \scleanbc (pink) samples. Also plotted are the Gaussian 
fits to these contours (green), see text for details.}\label{mcmc}
\end{figure}

We use MCMC analysis to explore our likelihood, (\ref{like}). As a consistency check
we confirm that we are able to recover various input parameters of (\ref{mod}) - although the 
partial sky coverage does give some degeneracies as expected, between $M$ and 
$D$ in particular. $D$ and $Q$ are constrained to be positive and we plot their 
marginalised constraints from our \cleanbc and \scleanbc samples in Figure~\ref{mcmc}. 
We have marginalised over all the other parameters, including the monopole component, and 
therefore have folded in any uncertainty that the bias effect and subsequent correction 
introduces.

We find that the constraints 
on $D$ and $Q$ are both consistent with 0, with upper 95\% limits of 
0.018 and 0.047 respectively from the \cleanbc sample, and 0.031 and 0.069 
from the \scleanbc sample. Alternatively, we fit Gaussians to the 
distributions and we find that the best fits means $\sim 0$ to 3 decimal 
places, with standard deviations for $D$ and $Q$ of 0.0084 and 0.014 
from the \cleanbc data, and 0.025 and 0.037 from the \scleanbc data. 
Due to the uncertainty of the monopole component, and the 
null dipole and quadrupole signal, we have no meaningful 
constraints on the direction of the vectors. 

We conclude that the Galaxy Zoo spin results are consistent with statistical 
isotropy. In the next section we compare this result to those of similar studies.



\subsection{Third-party Datasets}\label{third}

Previous studies of the spin of spiral galaxies include the recent work of
~\cite{Longo} (herein L07a), where the line of sight spin direction was determined 
by eye for $\sim2,800$ galaxies visually selected from the SDSS (DR5). 
An excess of S-wise 
galaxies was seen in the direction (RA,DEC)$\sim (202^\circ,25^\circ)$ 
(in Equatorial coordinates) indicating that the 
spin vectors predominantly align in the opposite direction 
(RA,DEC)$\sim (22^\circ,-25^\circ)$. The probability of the effect 
was estimated to be 0.2\% under the assumption 
of isotropy and binomial statistics, although Monte Carlo simulations were 
not performed.

Similar work was carried out earlier in~\cite{SZwise}, within the 
context of probing theories of galaxy formation. The spin dipole 
was evaluated for $\sim7,500$ galaxies and 
found to be in the direction (RA,DEC)=$(270^\circ,10^\circ)$, and 
(RA,DEC)=$(210^\circ,40^\circ)$ 
for a subsample of nearby galaxies ($|cz|<3000$km s$^{-1}$). 
However, the amplitude of the dipole was not found to be significant 
when compared to Monte Carlo simulations.

Curiously, the dipoles from 
these two analyses are in completely opposite directions. 
The samples cover different amounts and parts of the sky, 
with SDSS mainly in 
the Northern hemisphere and the sample of~\cite{SZwise} predominantly in 
the Southern hemisphere. In both cases the dipoles tend to point 
away from the majority of 
the data but neither analysis fits for a monopole or takes account of 
their partial sky coverage in assessing the dipole. 
With incomplete sky coverage the spherical harmonic decomposition is no 
longer orthogonal and for a sample covering less than half of the sky 
it is hard to tell the difference between a 
monopole (an excess of one type over the other) and a dipole (an 
asymmetry in the distribution).

\cite{Iye} compiled a catalog of the projected spins of 8,287 galaxies from the 
ESO/Uppsala Survey of the ESO (B) Atlas. Two independent judgments of the 
winding sense were made, and if these disagreed then a third judgement was 
taken. The result is a catalog of (3,257,  3,268)=(Z, S)-wise galaxies, with 
the remaining 1,762 galaxies deemed indistinguishable. These galaxies are 
all in the Southern hemisphere, and so provide a complementary dataset 
to ours which is primarily in the Northern hemisphere. 

We combine our \cleanbc dataset with their 6,525 galaxies with distinguishable handedness
and repeat our MCMC analysis fitting our probability model (\ref{mod}). 
However, in this case we must allow each dataset to have 
its own monopole term, to account for the different levels of possible bias in 
the results due to the different methods of classification. Again we find constraints 
that are completely consistent with 
zero, and we fit a Gaussian of $\sigma=0.0079$ and $0.017$ to the $D$ 
and $Q$ contours respectively. The constraints do not reduce as much as one would expect 
when expanding the dataset to the full sky because of the different monopole terms 
that we have included.

 \begin{table}
\begin{tabular}{lcccccccc}
\hline
{\bf Longo} & 1 & 2 & 3 & 4 & 5 & 6 & -  & tot \\
\hline
{\bf Z} & 0 & 1143 & 7 & 0 & 1 & 2 & 113 & 1266 \\
{\bf S} & 0 & 3 & 1235 & 0 & 1 & 1 &  125 & 1365 \\
{\bf U} & 0 & 43 & 63 & 0 & 1 & 1  & 95  & 223 \\
\hline
      & 0 & 1189 & 1305 & 0 & 3 & 4 & 333 & 2834 \\
\hline
\end{tabular}
\caption{The level of agreement between our \cleanbc classifications and 
those of L07a. The `-' column are those galaxies without a 
\cleanbc GZ classification.}\label{tabL}
\end{table}

We are able to do a more direct comparison with the dataset of L07a, as 
this also uses galaxies from SDSS and thus with which we overlap. 
The sample of L07a consists of 
$\sim 2,800$ galaxies from SDSS DR5 with redshifts $z<0.04$ and apparent 
magnitude constraint $g<17$ that were visually selected as being reasonably clear. 
Both computerised and visual methods of spin classification were attempted - and 
the results were found to agree well, but visually scanning 
was more effective towards higher redshifts - providing an extra $\sim 50\%$ 
galaxies. The classifications used in L07a are thus those from 
visual classification - where a galaxy image was randomly mirrored in the process. 

Of this sample, L07a found 1,266 and 1,365 galaxies to be Z and S-wise 
respectively, with the remaining `Unknown' (U)~\footnote{These numbers 
correspond to the dataset gratefully provided to the authors by M.Longo}. 
We find \cleanbc classifications for 2,501 of the sample, and 
the level of agreement is tabulated in Table~\ref{tabL}.
We see very good agreement between our classifications, and the 10 cases
where the handedness classifications contradict each other we have
double checked the SDSS images and find the disagreement can be put
down to misclassification in L07a - a
demonstration of the usefulness of multiple classifications per
galaxy.

If we fit just a dipole to the classifications of L07a using our formalism in
(\ref{mod}) with $Q=M=0$, then we find a clear detection of a dipole,
with constraints on $D$ well fit by a Gaussian with $D=0.035 \pm
0.017$ and best fitting direction $(RA, DEC)_{\hat{\bd}} = (-19^\circ,
-11^\circ)$.  This agrees with the signal detected by L07a;
a preference for galaxies to be S-wise in roughly the opposite direction
$(RA, DEC)\sim(202^\circ, 25^\circ)$. This was interpreted as a possible special
axis about which galaxies have a preferred handedness.

\begin{figure} 
\includegraphics[height=3.5cm]{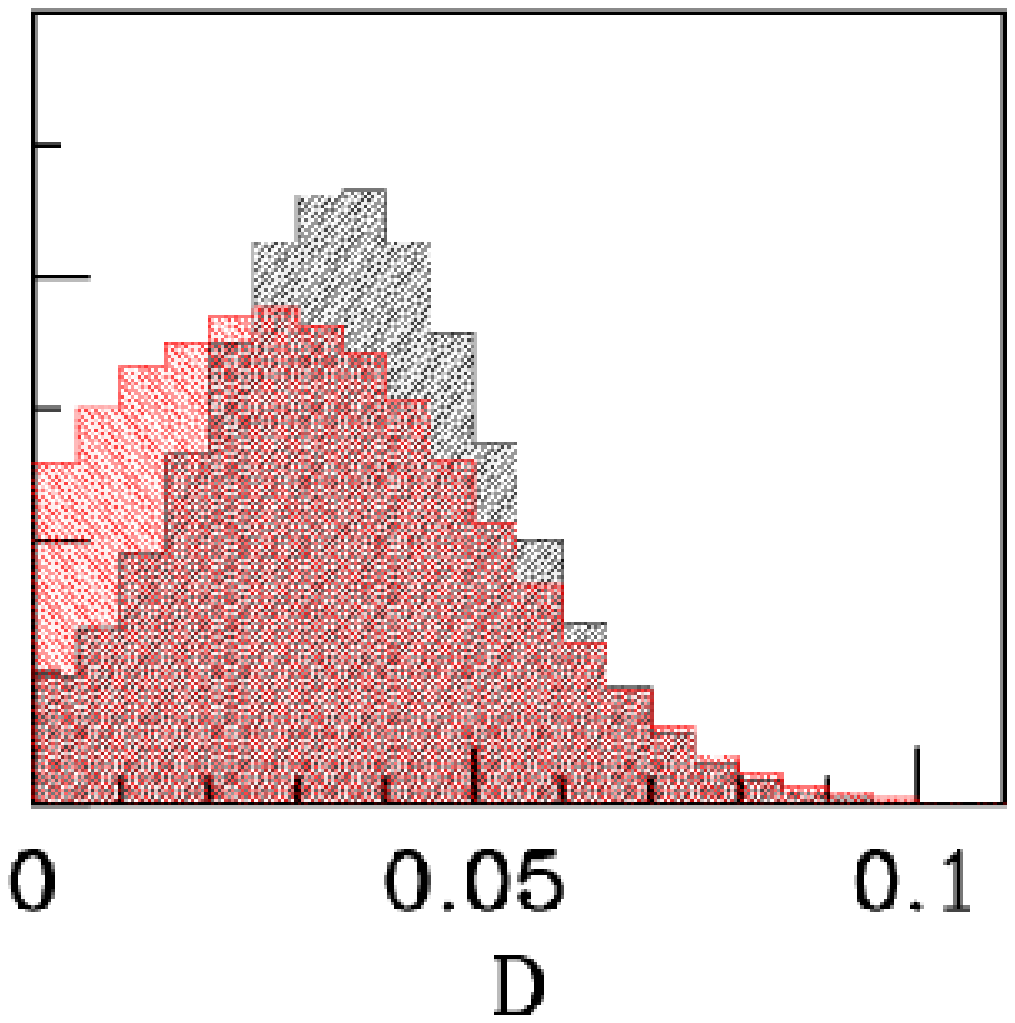}
\raisebox{0.15cm}{\includegraphics[height=3.4cm]{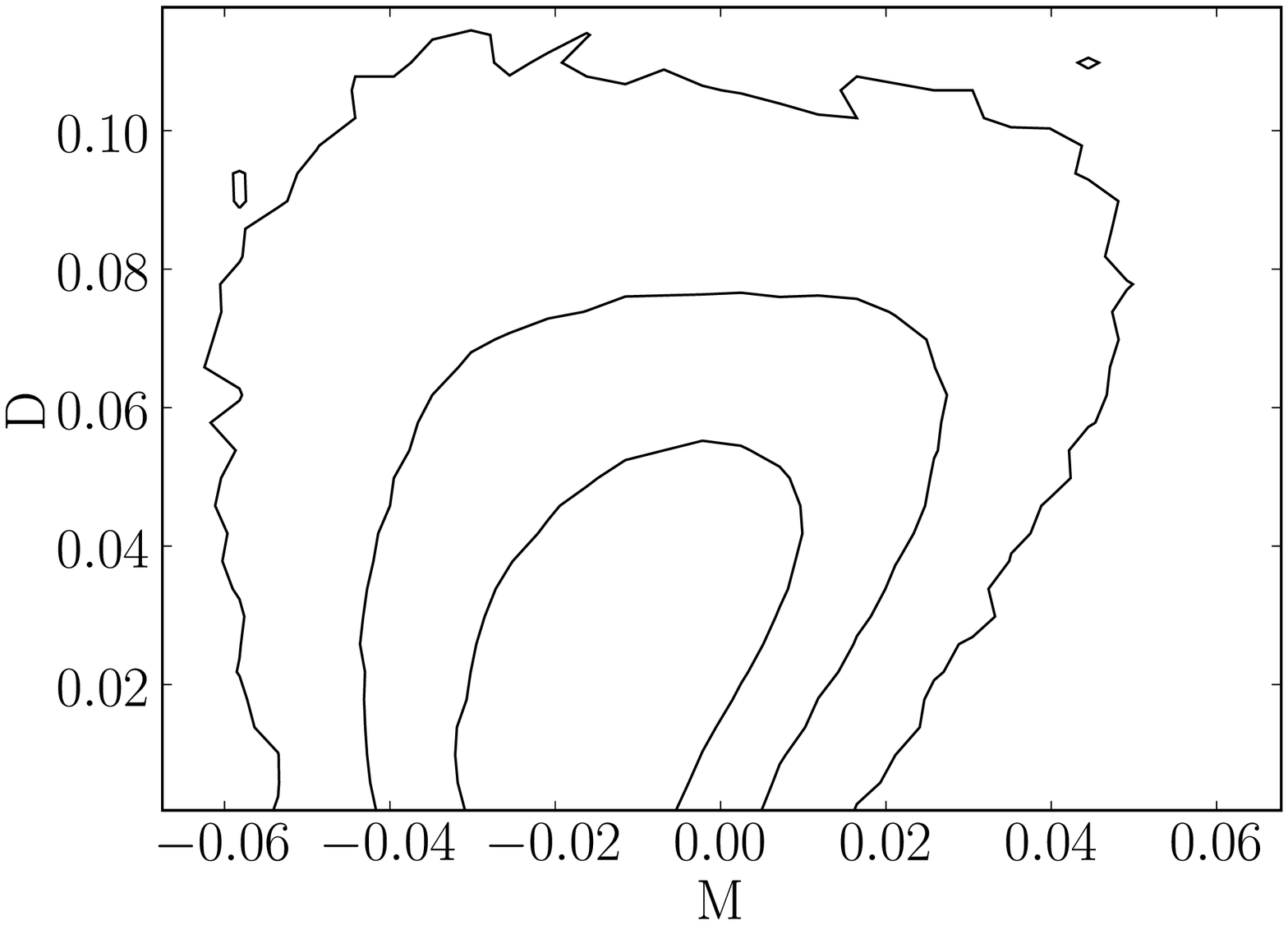}}
\caption{The constraints on the dipole from the sample of L07a, 
before (black histogram) and after (red) a monopole term is included and 
marginalised over. The right-hand figure shows the 1,2,3$\sigma$ 
contours for the monopole and dipole - demonstrating the degeneracy between the 
two as expected when you have only partial sky coverage.}\label{figL}
\end{figure}

However, with only partial sky coverage it is very hard to distinguish between 
a dipole and a monopole. And we repeat the analysis, but also 
marginalising over a monopole term (which can account for a true monopole
as well as any bias) in (\ref{mod}) but still excluding 
the quadrupole $Q$. We find the evidence for a dipole and thus for a preferred direction 
diminishes, and the constraints on $D$ become consistent with zero 
within $\sim 1\sigma$ as shown in Figure~\ref{figL}.

As seen in Table~\ref{tabL}, we agree that this subset of `visually clear' galaxies contains more S-wise 
galaxies than Z-wise (at $\sim 2 \sigma$), and that is after we have used our `bias corrected' 
\cleanbc catalogue of spins. However this alone does \emph{not} 
mean the L07a sample is clear of any bias effect: the galaxies in this sample were visually 
selected as being `clear' and judging by the results of 
Section~\ref{bias} this may explain why there are more S-wise galaxies are in the sample. 
A monopole term must always be fitted simultaneously to account for any level of bias in the 
dataset, and as we have seen this brings the observation in line with the hypothesis 
of statistical isotropy (i.e. no significant dipole).



\section{Conclusions}\label{end}

In this paper we have performed a classical test of cosmological 
statistical isotropy, from observations of the projected spins of galaxies. 
In line with the cosmological principle we find there is no significant 
large-scale correlations in the spin distributions (although a detection of 
the small-scale 2-point correlation function is found in~\cite{2point}). 

Due to the multiple classifications present in our dataset, we have 
been able to demonstrate that these visual classifications contain a certain 
level of bias between the identification of clockwise (Z-wise) and anti-clockwise (S-wise) 
rotating spiral galaxies. This must be accounted for in any analysis of the data, and is 
straightforward to do so by allowing the overall probability of the two (S vs Z) to 
differ. One then can continue to look for large-scale correlations such as a dipole, or a 
quadrupole.

We find that the projected angular momentum of galaxies in our local Universe are 
consistent with isotropy. This is in agreement with other similar studies~\citep{SZwise}, 
although our sample is a factor of 5 times larger than any previous datasets.
Our classifications agree very well with those of~\cite{Longo}, who found evidence for a 
preferred axis in their sample. We find that this evidence diminishes when one allows for an 
overall monopole term, to account for the unknown level of bias in the results (or a true 
signal on the sky).

These results aid our understanding of the formation of large-scale structure. 
They limit the possibility of coherent large-scale
magnetic fields that assist in the formation of galactic angular 
momentum~\citep{Longomag}. 
Similarly they constrain scenarios of galaxy formation - providing support for 
tidal torque theory~\citep{ttt} as opposed to scenarios in which the angular momentum of 
galaxies are coherent over large scales. Such scenarios include those where 
the angular momentum of galaxies originates from primordial vorticity, or from 
collapsing inflows~\citep{SZwise}.


\section*{Acknowledgements}
We thank M.Longo for many useful conversations, and for kindly sharing his 
dataset with us. We are similarly grateful to H.Sugai and M.Iye for assistance, particularly in 
recovering and sharing their dataset. We are grateful to P.Ferreira and J.Silk 
for many useful conversations and encouragement.

But ultimate thanks must 
of course go to all the volunteers that have made 
Galaxy Zoo possible. Not only does this include $\sim100,000$ users who personally 
identified hundreds of galaxies, but also 
web-designers, programmers, forum\footnote{www.galaxyzooforum.org} moderators, bloggers, journalists, 
guinea-pigs, colleagues, school teachers, etc. everyone who has been 
interested and excited, about the project! In particular we acknowledge invaluable 
contributions from Edd Edmondson and Alice Sheppard.

KL is funded by a Glasstone research fellowship, 
and further supported by Christ Church, Oxford. CJL acknowl-
edges support from the STFC Science in Society Program. KS is supported
by the Henry Skynner Junior Research Fellowship at Balliol College,
Oxford.


\bibliographystyle{mn2e_eprint}
\bibliography{GZ_Land}

\label{lastpage}

\end{document}